\documentclass{article}

\PassOptionsToPackage{numbers}{natbib}
\usepackage[preprint]{neurips_2021}

\usepackage[utf8]{inputenc} 
\usepackage[T1]{fontenc}    
\usepackage{hyperref}        
\usepackage{url}            
\usepackage{booktabs}       
\usepackage{amsfonts}       
\usepackage{nicefrac}       
\usepackage{microtype}      
\usepackage{xcolor}         
\usepackage{graphicx}
\usepackage{multirow}
\usepackage{float}
\usepackage{amsmath}

\title{Trajectory Inference of Human Aging from Cross-Sectional DNA Methylation Data}

\author{
Chandan Gupta \\
Independent Researcher \\
\\
\texttt{chandan18386@iiitd.ac.in} \\
\And
Syed Haider \\
The Institute of Cancer Research \\
United Kingdom \\
\texttt{syed.haider@icr.ac.uk} \\
\And
Pietro Liò \\
University of Cambridge \\
United Kingdom \\
\texttt{pl219@cam.ac.uk} \\
}

\begin{document}

\newcommand{\cpgpt}{CpGPT}
\newcommand{\methylgpt}{MethylGPT}
\newcommand{\altumage}{\textbf{Altumage}}
\newcommand{\preten}{\textbf{Pretrain-10}}
\newcommand{\pretwen}{\textbf{Pretrain-20}}

\maketitle

\begin{abstract}
DNA methylation (DNAm) serves as one of the most robust molecular biomarkers of biological aging. While conventional epigenetic clocks accurately predict chronological age from high-dimensional CpG profiles, they treat aging as a static regression task, meaning they can only output a single score rather than simulating how an entire profile continuously changes over time. To reconstruct these continuous dynamics, we frame lifelong human epigenetic aging as a trajectory inference problem across discrete age snapshots derived from widely available cross-sectional data. We introduce a two-stage computational pipeline: first, an age-regularized Variational Autoencoder (VAE) maps high-dimensional CpG profiles onto a chronologically ordered latent manifold while preserving a generative decoder bridge back to the original methylation space. Second, we model the continuous movement across this latent space via Regularized Unbalanced Optimal Transport (RUOT) that unifies deterministic drift, random diffusion, and non-conservative mass changes. By resolving this RUOT formulation using the DeepRUOT framework, our model fluidly accommodates population-level density shifts like survivorship bias and cellular attrition without requiring rigid biological priors. Evaluated on a large-scale, 80-year pan-tissue dataset, our model demonstrates robust distribution interpolation and uncovers a prominent late-life surge in the learned growth field that mathematically captures the variance expansion driven by stochastic epigenetic drift. Finally, by decoding continuous latent paths back to individual CpG sites, we reconstruct and empirically verify distinct biological aging archetypes, offering a rigorous, generative paradigm for simulating human molecular aging.
\end{abstract}

\section{Introduction}

In computational biology, reconstructing continuous stochastic dynamics from static, cross-sectional snapshots is a central challenge. Because many high-throughput assays are destructive, true longitudinal trajectories are rarely observed. While this problem has been extensively studied in single-cell transcriptomics, where trajectory inference and dynamic optimal transport (OT) models successfully reconstruct continuous developmental continuums from unlinked cell populations \cite{schiebinger2019optimal, bunne2023learning, lavenant2021towards, bunne2024optimal}, extending these stochastic models to map the continuous, lifespan-wide progression of human biological aging remains largely unaddressed.

Currently, DNA methylation (DNAm) serves as a robust biomarker of biological aging \cite{jones2015dna, field2018dna}. Supervised "epigenetic clocks" have revolutionized this domain by predicting chronological age from high-dimensional CpG profiles \cite{horvath2013dna, hannum2013genome, levine2018epigenetic, lu2019dna}. However, by compressing profiles into a scalar point estimate, they treat aging as a static regression task and lack the generative capacity to simulate how the joint probability distributions of the epigenetic landscape evolve over time. Transitioning from scalar tracking to a continuous dynamical system is essential to uncover these joint distribution flows and capture the true mechanics of molecular senescence.

Recent frameworks have approached DNAm kinetics through specialized lenses, ranging from univariate curve-fitting \cite{snir2019human} and fractional calculus \cite{nasrolahpour2025fractional} to macroscopic parametric SDEs \cite{allenbiological, zagkos2021mathematical}. However, these approaches fail to scale to organismal aging: they either focus on isolated variables that miss global co-dependencies, or rely on rigid, mass-conserving parametric forms. Crucially, population-level aging is an unbalanced, non-conservative system. Phenomena such as demographic survivorship bias, where fast-aging individuals are selectively removed via early mortality \cite{marioni2015dna}, and tissue-specific cellular attrition \cite{jaffe2014accounting} violently violate mass conservation. Traditional mass-preserving OT frameworks \cite{benamou2000computational, leonard2013survey} are thus structurally incapable of reconciling younger, dense cohorts with older, scattered survivors without artificially distorting the underlying trajectories.

To address this, we formulate lifelong epigenetic aging as a trajectory inference problem across continuous time by leveraging static, cross-sectional datasets. Specifically, we discretize the population into chronological age bins that serve as sequential, unlinked temporal snapshots of the evolving methylome distribution. We then resolve the non-conservative transitions between these age snapshots by reformulating the dynamics as a Regularized Unbalanced Optimal Transport (RUOT) problem \cite{chen2022most}. Also known as the unbalanced Schrödinger Bridge, RUOT mathematically unifies stochastic diffusion with mass-flexibility fields \cite{chizat2018interpolating}. We track these kinetics using the DeepRUOT solver \cite{zhang2025learning}, which uses the snapshot distributions to learn three coupled components: a velocity field ($v_\theta$) for directional aging drift, a score function ($s_\theta$) for random noise, and an independent growth field ($g_\theta$) that acts as a mathematical pressure valve to adjust population density as variance expands in later life. To bridge the gap between high-dimensional arrays and optimal transport, we construct an age-regularized latent manifold that isolates clean aging signals from tissue heterogeneity. Our main contributions are:

\begin{itemize}
    \item \textbf{Age-Aware Latent Architecture:} We introduce a novel age-regularized Variational Autoencoder (VAE) \cite{Kingma2014} that projects bulk DNAm profiles into a chronologically ordered latent space while maintaining a generative decoder bridge back to the original CpG feature space.
    \item \textbf{Application of Unbalanced Trajectory Inference:} We deploy the DeepRUOT framework across this aging manifold, demonstrating that its unbalanced formulation naturally captures the localized mass adjustments required to simulate late-life entropic expansion and survivorship bias across the discrete age cohorts.
    \item \textbf{Empirical Verification of Aging Archetypes:} We validate the pipeline on a large-scale pan-tissue dataset, decoding continuous latent paths to successfully reconstruct and empirically verify distinct biological archetypes: linear clocks, stable maintenance sites, and accelerated late-onset drift.
\end{itemize}

\section{Methodology}

\begin{figure}
    \centering
    \includegraphics[width=\linewidth]{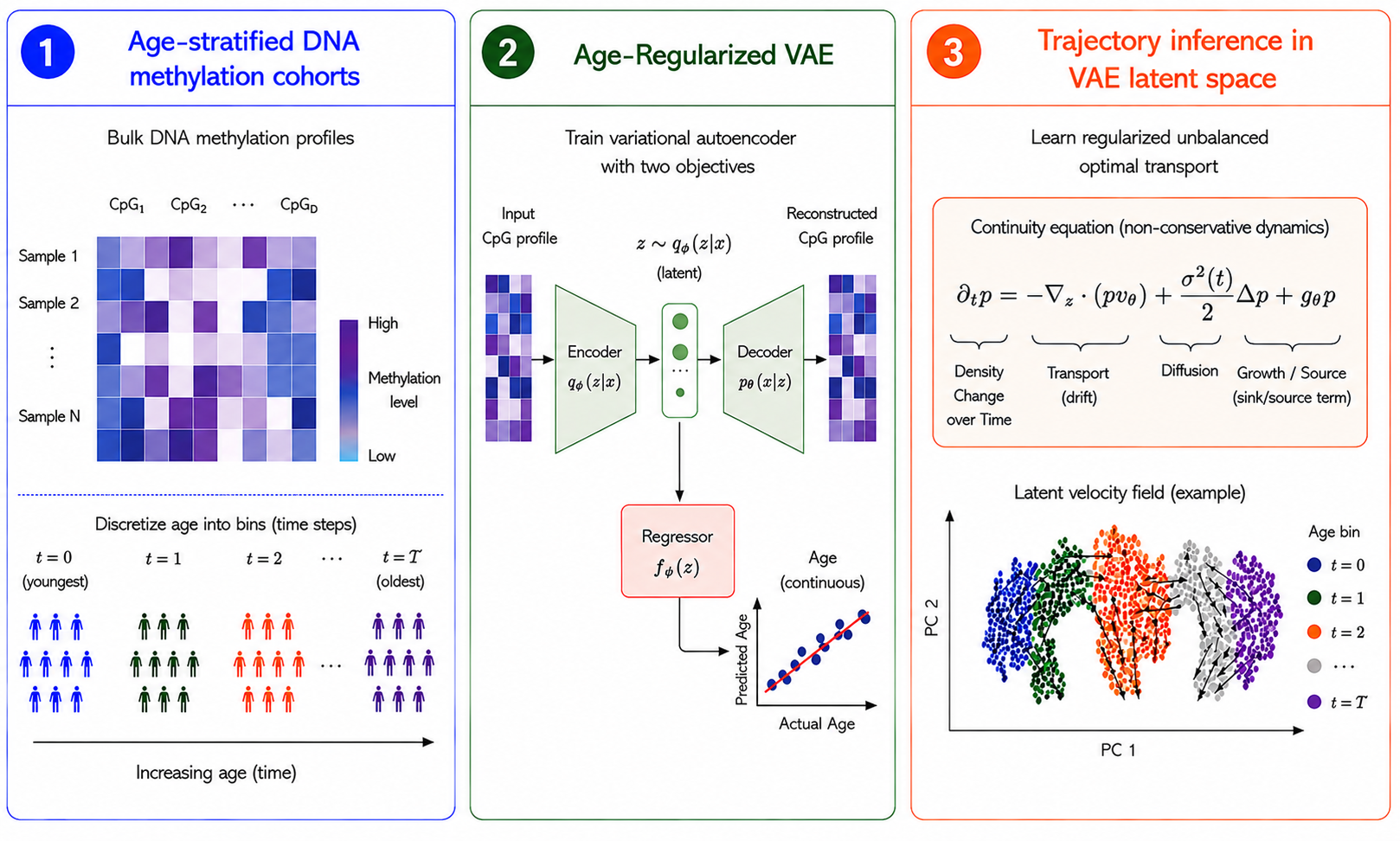}
    \caption{Methodological framework for inferring continuous biological aging trajectories. (1) Cross-sectional DNA methylation snapshots are discretized into age-based cohorts. (2) An age-regularized Variational Autoencoder (VAE) compresses these high-dimensional profiles into a biologically structured latent space that is inherently age-aware, ensuring chronological progression is preserved. (3) Latent aging dynamics are modeled using regularized optimal transport, inferring the continuous velocity and growth fields that describe population-level epigenetic drift across the aging manifold.}
    \label{fig:methodology}
\end{figure}

To resolve the extreme temporal ambiguity and non-conservative distribution shifts inherent in cross-sectional epigenetic data, we propose a two-stage computational pipeline that couples latent manifold structuring with continuous dynamic inference (Figure \ref{fig:methodology}). Rather than optimizing optimal transport directly on high-dimensional, noisy CpG arrays, our framework operates sequentially: it first maps cross-sectional profiles into a compressed, chronologically ordered latent space, and subsequently solves a parameter-free regularized unbalanced optimal transport problem across this age-aware manifold to reconstruct continuous lifespan-wide aging trajectories.

We describe each component in detail below.

\subsection{Age-Regularized Variational Autoencoder}To project high-dimensional epigenetic profiles $x \in \mathbb{R}^D$ into a computationally tractable latent space $\mathcal{Z} \subset \mathbb{R}^d$, we employ a supervised Variational Autoencoder architecture. Standard VAEs optimize the Evidence Lower Bound (ELBO), comprising a reconstruction loss ($\mathcal{L}_{\text{recon}}$) and a Kullback-Leibler divergence ($\mathcal{L}_{\text{KL}}$) that regularizes the latent distribution toward a standard normal prior \cite{Kingma2014}. To explicitly enforce chronological organization within this geometry, we append an auxiliary age-regression network $f_\psi(z)$ directly to the latent bottleneck. The joint architecture—comprising an encoder $q_\phi(z|x)$, a generative decoder $p_\theta(x|z)$, and the regressor $f_\psi(z)$—is optimized via a composite loss function:
\begin{equation}\mathcal{L}_{\text{VAE}} = \mathcal{L}_{\text{recon}} + \beta_{KL} \mathcal{L}_{\text{KL}} + \gamma \mathcal{L}_{\text{age}}\end{equation}

where $\mathcal{L}_{\text{age}}$ represents the Mean Squared Error (MSE) of the predicted chronological age derived from the latent vector $z$. The weighting hyperparameters $\beta_{KL}$ and $\gamma$ dictate the topological constraints of the manifold. Following the $\beta$-VAE framework \cite{higgins2017beta}, the penalty $\beta_{KL}$ enforces smoothness and structural disentanglement within the latent distribution, while the age penalty $\gamma$ forces this latent space to align monotonically with biological time. This joint regularization ensures that our discrete cross-sectional cohorts form a continuous, age-aware progression suitable for downstream optimal transport modeling. Crucially, the preserved generative decoder $p_\theta(x|z)$ enables bidirectional mapping: any continuous transport trajectory simulated within $\mathcal{Z}$ can be decoded back into the original CpG feature space, transforming abstract latent flows into interpretable, site-specific biomarker curves. 

\subsection{Trajectory Inference via DeepRUOT}

To infer continuous aging dynamics from the discrete, age-binned latent snapshots, we utilize the Regularized Unbalanced Optimal Transport (RUOT) \cite{chen2022most} framework via its deep learning solver, DeepRUOT \cite{zhang2025learning}. Standard optimal transport assumes strict mass conservation. However, population-level epigenetic aging is a non-conservative system subject to runtime noise, survivorship bias, and cellular attrition. RUOT accommodates these phenomena by unifying the stochastic diffusion properties of the Schrödinger Bridge problem \cite{leonard2013survey} with the unnormalized mass flexibility of unbalanced dynamic optimal transport governed by the Wasserstein–Fisher–Rao (WFR) metric \cite{chizat2018interpolating}.

Formally, given initial and terminal latent mass densities $\nu_0$ and $\nu_1$, RUOT minimizes the joint kinetic energy and mass-alteration effort under a generalized Fokker-Planck constraint:
\begin{align}
    \inf_{(p,b,g)} \quad &\int_0^1 \int_{\mathbb{R}^d} \left[ \frac{1}{2} \|b(z, t)\|^2_2 + \alpha \Psi(g(z, t)) \right] p(z, t) \, dz \, dt \\
    \text{s.t.} \quad &\partial_t p = -\nabla_z \cdot (pb) + \frac{\sigma^2(t)}{2} \Delta p + gp, \quad p(\cdot, 0) = \nu_0, \quad p(\cdot, 1) = \nu_1 \nonumber
\end{align}
where $b(z,t)$ represents the drift vector field, $g(z,t)$ is a scalar growth/decay rate field, $\sigma(t)$ dictates the diffusion rate, and $\Psi(\cdot)$ is a quadratic growth penalty function ($\Psi(g) = \frac{1}{2}|g|^2$).

DeepRUOT overcomes the high-dimensional intractability of this formulation by applying a Fisher information regularization form \cite{zhang2025learning}. This converts the underlying SDE system into an equivalent deterministic path-mapping system parameterized by three independent neural networks: a velocity field $v_{\theta_v}(z, t)$, a growth field $g_{\theta_g}(z, t)$, and a scaled log-density score network $s_{\theta_s}(z, t) = \frac{\sigma^2(t)}{2} \log p(z, t)$. The entire system is trained end-to-end by minimizing a composite loss function over pseudo-time $t \in [0, T]$ across $K$ discrete age cohorts:
\begin{equation}
    \mathcal{L}_{\text{RUOT}} = \mathcal{L}_{\text{Energy}} + \lambda_r \mathcal{L}_{\text{Recons}}
\end{equation}

\paragraph{Energy Loss} Utilizing Monte Carlo sampling and a neural ODE solver along the characteristic paths $\frac{dz}{dt} = v_{\theta_v}(z, t)$, the dynamic energy penalty derived from the Fisher regularization form is optimized via:
\begin{equation}
\begin{split}
    \mathcal{L}_{\text{Energy}} = \mathbb{E}_{z_0 \sim p_0} \int_0^T \bigg[ &\frac{1}{2} \|v_{\theta_v}\|^2_2 + \frac{1}{2} \|\nabla_z s_{\theta_s}\|^2_2 - \left(\frac{\sigma^2(t)}{2} + s_{\theta_s}\right)g_{\theta_g} \\
    &- \frac{(\sigma^2(t))'}{\sigma^2(t)}s_{\theta_s} + \alpha \Psi(g_{\theta_g}) \bigg] w_{\theta_g}(t) \, dt
\end{split}
\end{equation}
where $w_{\theta_g}(t) = \exp\left(\int_0^t g_{\theta_g}(z(s), s) \, ds\right)$ acts as a dynamic particle weight tracker.

\paragraph{Reconstruction Loss} To align the continuous trajectories with unnormalized empirical distributions at subsequent age checkframes, the reconstruction loss unifies local mass matching and normalized transport:
\begin{equation}
    \mathcal{L}_{\text{Recons}} = \lambda_m \mathcal{L}_{\text{Mass}} + \lambda_d \mathcal{L}_{\text{OT}}
\end{equation}
Here, $\mathcal{L}_{\text{Mass}} = \sum_{k=1}^{K-1} M_k$ computes a local mass matching error using a localized cardinality mapping $h_k$ between observed cohort data points $A_k$ and their nearest simulated counterparts $\hat{A}_k$. Once particle weights $w_i(t_k)$ satisfy local boundary criteria, $\mathcal{L}_{\text{OT}}$ minimizes the standard $W_2$ Wasserstein distance between the normalized empirical cohort distributions and the weight-adjusted simulated distributions.

\section{Experimental Setup}

\subsection{Dataset Description}
We utilized the Altumage dataset \cite{de2022pan}, a large-scale collection of publicly available DNA methylation profiles. Rather than restricting our analysis to a single tissue, we retained a diverse, pan-tissue cohort encompassing whole blood, brain, saliva, and various solid tissues. This pan-tissue approach, allows us to infer a universal, systemic trajectory of chronological aging that is robust to tissue-specific epigenetic signatures.

\subsection{Data Preprocessing}
The filtered dataset was partitioned into training (70
To reduce dimensionality while retaining highly age-associated features, we performed univariate feature selection using the ANOVA F-statistic. We retained the top 2,000 CpG sites most strongly correlated with chronological age in the training set. To ensure statistical robustness during trajectory inference, we restricted the target population to the age range of 0 to 80 years, resulting in a final cohort of 12,439 valid samples. Individuals were discretized into four chronological age bins: 0--20 years ($t_0$), 21--40 years ($t_1$), 41--60 years ($t_2$), and 61--80 years ($t_3$), which served as the temporal anchors for the DeepRUOT framework. 

\subsection{Model Architectures and Training}

The age-regularized VAE uses a symmetric Multi-Layer Perceptron (MLP) architecture where the encoder compresses the 2,000-dimensional input through hidden layers of 512 and 256 units into a 16-dimensional latent space ($d=16$). The decoder mirrors this structure to reconstruct $M$-values, while an auxiliary regression head ($16 \to 64 \to 1$) predicts chronological age from the latent bottleneck. All hidden layers incorporate Batch Normalization, LeakyReLU activations, and dropout (rate = 0.2) \cite{srivastava2014dropout}. The unified network is optimized via Adam (learning rate = $1 \times 10^{-3}$, weight decay = $1 \times 10^{-5}$) \cite{kingma2014adam} with a batch size of 128 to minimize $\mathcal{L}_{\text{VAE}}$ ($\beta_{KL} = 0.1, \gamma = 0.3$), employing early stopping with a 25-epoch patience on the validation set.

For trajectory inference, the DeepRUOT velocity $v_{\theta_v}$ and growth $g_{\theta_g}$ fields are parameterized by independent 2-layer, 400-unit MLPs with LeakyReLU activations, while the score network $s_{\theta_s}$ utilizes a 128-unit hidden dimension. Optimization proceeds via Adam (learning rate = $1 \times 10^{-4}$) using bootstrap sampling with replacement ($N=1,000$) across three distinct phases, keeping the physics-informed constraint deactivated ($\text{use\_pinn}=\text{false}$). First, a 500-epoch pre-training phase optimizes mass reconstruction ($\lambda_{\text{mass}} = 0.01, \lambda_{\text{OT}} = 1.0, \lambda_{\text{energy}} = 0.0$). Second, the score network $s_{\theta_s}$ is trained independently for 3,001 epochs (batch size = 128, $\sigma = 0.1, \lambda_{\text{penalty}} = 1$). Finally, the complete loss $\mathcal{L}_{\text{RUOT}}$ is minimized end-to-end for 500 epochs ($\lambda_{\text{energy}} = 0.01, \lambda_{\text{mass}} = 10, \lambda_{\text{OT}} = 10$) using a step learning rate scheduler (step size = 100 epochs, decay $\gamma_{\text{sched}} = 0.8$).

\textbf{Baseline : }To evaluate the necessity of our unbalanced trajectory inference formulation, we established a baseline using standard mass-conserving optimal transport (OT). Rather than utilizing a separate architecture, we simulated this baseline directly within our framework by systematically disabling the unbalanced growth field ($g_\theta = 0$) during optimization. This ablation isolates the exact contribution of the growth term, allowing us to quantitatively assess its role in modelling non-conservative population dynamics, such as survivorship bias and demographic mass destruction, compared to rigid, mass-conserving methods.

\section{Results}
\subsection{Quantitative Evaluation}

To quantitatively evaluate the framework's consistency, we performed a leave-one-timepoint-out cross-validation, measuring the Wasserstein-1 ($W_1$) distance and Total Mass Variation (TMV) between the predicted and empirical target distributions. Intuitively, these two metrics capture distinct aspects of the aging dynamics. The $W_1$ distance evaluates the overall structural accuracy of the prediction, measuring how closely the model's generated epigenetic profiles geometrically align with the actual patient profiles. TMV, conversely, quantifies the amount of localized density adjustment required by the DeepRUOT. Biologically, TMV acts as a mathematical proxy for population-level divergence, tracking how much the variance of a cohort expands or disperses due to biological noise.

As shown in Table~\ref{tab:altumage_results}, the model yields relatively stable $W_1$ distances across all hold-out experiments, indicating that the learned vector fields can smoothly interpolate cross-sectional epigenetic distributions across unobserved temporal gaps without catastrophic divergence. Crucially, while $W_1$ remains stable, the TMV reveals a distinct shift in aging dynamics, as it is minimal during early-to-mid-life transitions ($0.754$ for $T=1$; $0.620$ for $T=2$). However, predicting the oldest age bin requires a substantial, nearly four-fold increase in TMV ($2.321$ for $T=3$). 

\begin{table}[H]
\centering
\caption{Quantitative baseline comparison of trajectory inference robustness. DeepRUOT (with unbalanced growth) is compared against a Standard OT baseline (mass-conserving, $g_\theta=0$) across held-out age bins.}
\label{tab:altumage_results}
\begin{tabular}{llccc}
\toprule
\textbf{Model Configuration} & \textbf{Metric} & \textbf{T = 1} & \textbf{T = 2} & \textbf{T = 3} \\
\midrule
\multirow{2}{*}{\textbf{DeepRUOT (With Growth)}} 
& $W_1$ & 4.379 $\pm$ 0.001 & 3.574 $\pm$ 0.006 & 3.824 $\pm$ 0.025 \\
& TMV   & 0.754 $\pm$ 0.000 & 0.620 $\pm$ 0.000 & 2.321 $\pm$ 0.012 \\
\midrule
\multirow{2}{*}{\textbf{Standard OT ($g_\theta = 0$)}} 
& $W_1$ & 3.996 $\pm$ 0.002 & 4.453 $\pm$ 0.004 & 6.770 $\pm$ 0.007 \\
& TMV   & 0.394 $\pm$ 0.000 & 0.200 $\pm$ 0.000 & 0.076 $\pm$ 0.000 \\
\bottomrule
\end{tabular}
\end{table}

As introduced in our baseline setup, we benchmarked DeepRUOT against standard optimal transport ($g_\theta = 0$). Standard OT mathematically enforces mass conservation, incorrectly driving TMV to near zero (e.g., $0.076$ at $T=3$). Without the growth field to handle non-conservative biological realities like survivorship bias and demographic mass destruction, standard OT fails to geometrically align with the late-life data, resulting in a substantially higher $W_1$ error ($6.770$ vs DeepRUOT's $3.824$). DeepRUOT’s higher TMV ($2.321$) is thus mathematically required to accurately capture this late-life variance expansion.

\begin{figure}
    \centering
    \includegraphics[width=0.9\linewidth]{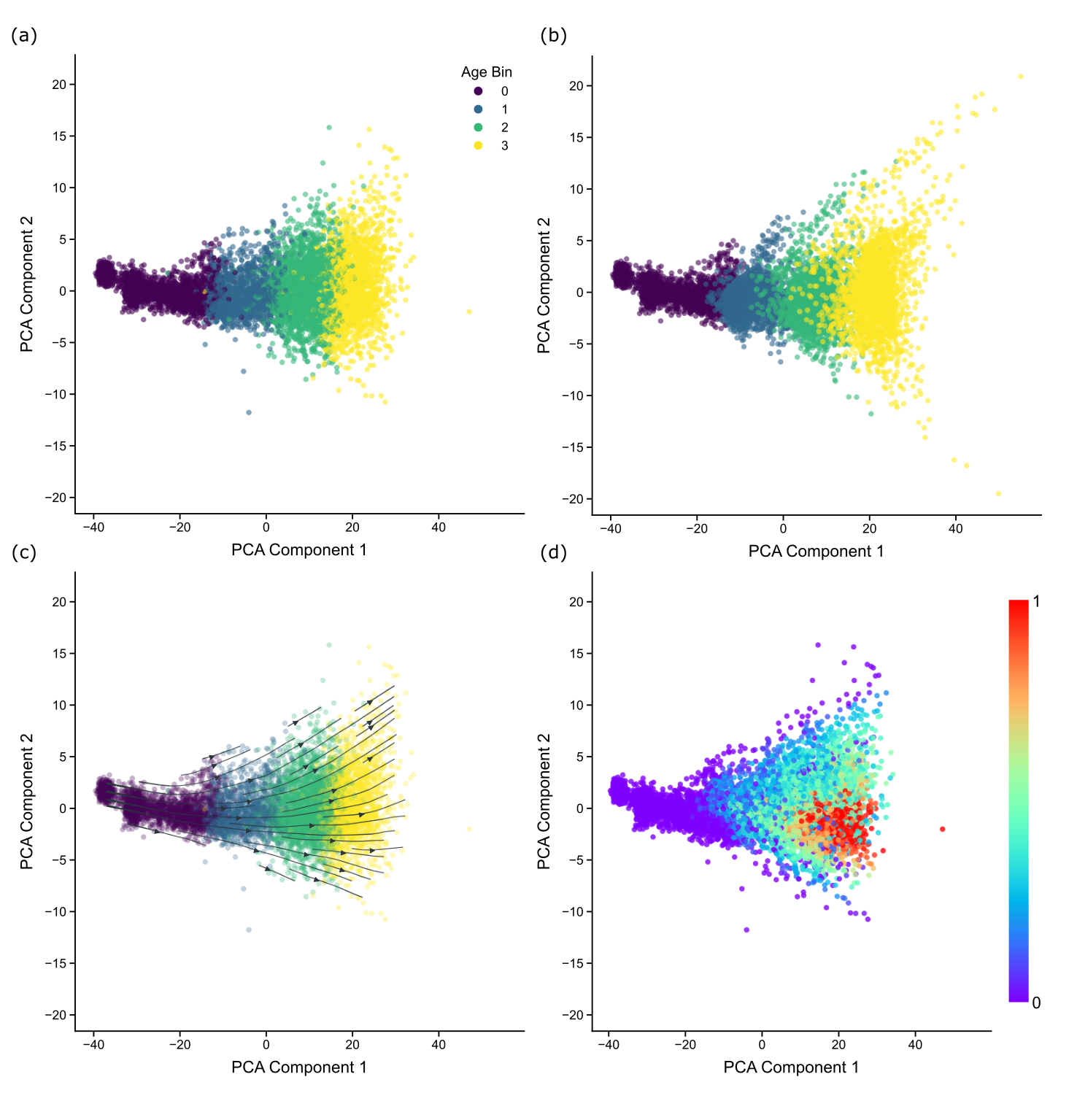}
    \caption{Trajectory inference of human epigenetic aging in latent space. (a) Ground truth PCA projection. (b) Predicted trajectories PCA projection. (c) The inferred velocity field streamlines showing continuous chronological progression. (d) Normalized learned growth rates showing localized mass adjustments.}
    \label{fig:velocity}
\end{figure}

We then visualised the the generated trajectories from the model. Figure~\ref{fig:velocity}(a) and Figure~\ref{fig:velocity}(b) illustrate that the model successfully reconstructs the empirical data manifold, closely matching the expanding spatial variance observed in the late-life distributions. The continuous velocity field visualized in Figure~\ref{fig:velocity}(c) reveals a globally stable, unidirectional flow from youthful baseline states (purple) toward late-life states (yellow). The inferred streamlines smoothly map the continuous chronological progression across the latent manifold.

While the continuous velocity field tracks the overall direction of the aging trajectory (Figure~\ref{fig:velocity}c), the growth term reveals where the population distribution undergoes rapid, unpredictable expansion. As shown in Figure~\ref{fig:velocity}(d), the predicted growth rate remains near zero throughout early and middle life. This indicates a period of stable, predictable biological progression where individuals age along tightly bundled paths without extreme deviation. However, as trajectories progress into late-life states, the model exhibits a prominent, localized surge in the growth term. Rather than forcing the older population into a narrow, unrealistic pathway, this localized growth injects the necessary variance into the system, allowing the simulated trajectories to naturally spread out.

We hypothesize that this mathematical shift mirrors the biological onset of stochastic epigenetic drift. As established by previous studies \cite{fraga2005epigenetic, bertucci2023rate, seale2022making, slieker2016age}, chronological aging causes random methylation errors to accumulate, making older populations much more diverse than younger ones. This loss of precision is driven by the gradual breakdown of cellular maintenance systems \cite{lopez2023hallmarks}. Because older individuals become so biologically scattered, standard models that force everyone along a strict, uniform aging path end up distorting the data. By relaxing these rigid constraints, the growth term allows the simulated trajectories to fan out, naturally accommodating the messy and unpredictable nature of late-stage aging.

\subsection{Biological Characterization of Epigenetic Archetypes}
To reconstruct continuous chronological aging trajectories from our discrete age bins, we first extracted the mean latent path. We linearly mapped this progression to an 80-year chronological timeline, yielding a continuous latent trajectory $z(t)$. We then projected $z(t)$ back through the age-regularized VAE decoder to generate corresponding DNAm profiles, producing a continuous, 80-year aging curve $\beta_i(t)$ for every individual CpG site.

\begin{figure}
    \centering
    \includegraphics[width=0.9\linewidth]{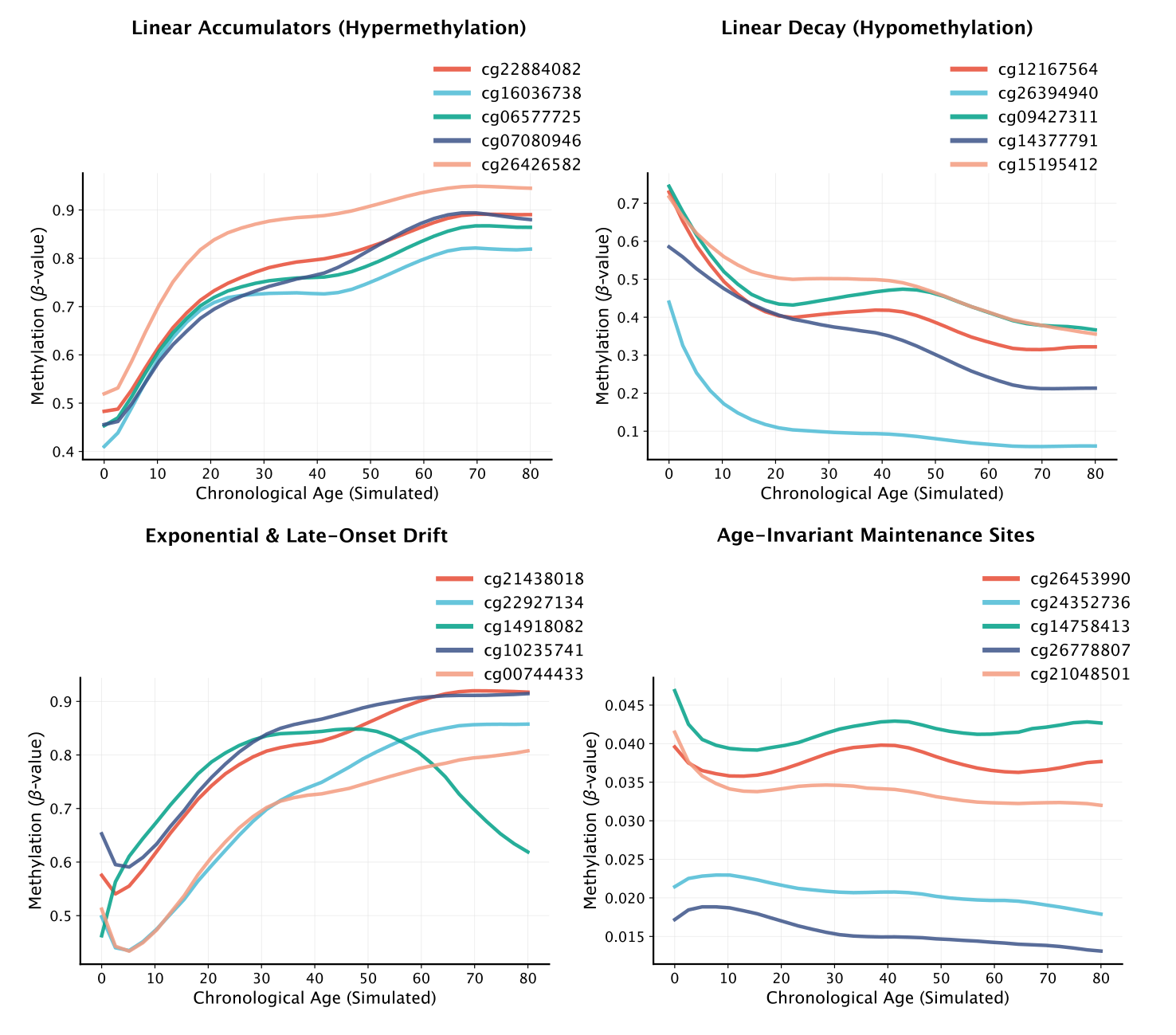}
    \caption{Biological archetypes of DNA methylation aging. Continuous trajectories were generated by decoding the inferred latent flow back into the original CpG feature space. (Top-Left) Linear hypermethylation. (Top-Right) Linear hypomethylation. (Bottom-Left) Accelerated late-onset stochastic drift. (Bottom-Right) Age-invariant maintenance loci.}
    \label{fig:archetypes}
\end{figure}

To automatically categorize these individual biomarkers into distinct biological archetypes, we analyzed the kinetic properties of each curve $\beta_i(t)$ across the simulated lifespan. We define four primary aging archetypes (Figure \ref{fig:archetypes}):

\begin{itemize}
    \item \textbf{Linear Accumulators (Hypermethylation):} Defined by the maximum positive shift over the lifespan, $\max (\beta_i(80) - \beta_i(0))$. These sites exhibit a monotonic, steady gain in methylation. They predominantly reside in the promoters of developmental genes \cite{rakyan2010human}, representing the "clock-like" precision of the epigenetic system as it systematically silences developmentally regulated loci \cite{horvath2013dna}.
    
    \item \textbf{Linear Decay (Hypomethylation):} Defined by the maximum negative shift over the lifespan, $\min (\beta_i(80) - \beta_i(0)).$ These markers display a gradual, global loss of methylation. This archetype is highly characteristic of repetitive genomic elements (e.g., Alu and LINE-1) \cite{bollati2009decline}, reflecting the cumulative erosion of structural integrity and the progressive loss of maintenance methyltransferase fidelity \cite{ cruickshanks2013senescent}.
    
\item \textbf{Exponential \& Late-Onset Drift:} Defined by the maximum absolute curvature (second derivative) across the trajectory:
    \begin{equation}
        \max_t \left| \frac{d^2 \beta_i}{dt^2} \right| \approx \max_k \left| \beta_i(t_{k+1}) - 2\beta_i(t_k) + \beta_i(t_{k-1}) \right|
    \end{equation}
    This class captures markers that remain stable during early and mid-life but undergo sudden, accelerated divergence. These sites are key molecular indicators of the mathematical inflection point identified in our growth-rate analysis, serving as a sensitive proxy for the onset of late-life biological entropy and the non-linear accrual of methylomic variability \cite{slieker2016age, snir2019human}.

    \item \textbf{Age-Invariant Maintenance Sites:} Defined as the minimum temporal variance across the trajectory, $\min \text{Var}(\beta_i(t))$. These loci maintain a constant methylation status throughout the entire lifespan \cite{ziller2013charting, issa2014aging}. They represent the static, robust "housekeeping" component of the epigenome, which is strictly preserved to maintain essential cellular functions and heavily protected against the stochastic noise that otherwise drives epigenetic drift in unprotected regions. 

\end{itemize}

The divergence in these archetypes highlights that epigenetic aging is not a uniform process, but a composite of distinct kinetic behaviors. By successfully reconstructing these distinct profiles from a single global velocity field, our model confirms that population-level epigenetic drift is a highly localized, site-specific phenomenon optimally captured by the framework. 

We further verify the continuous paths generated in the latent space by overlaying the decoded mean trajectories directly against the raw, cross-sectional data. Across all archetypes, the inferred paths faithfully track the moving center of mass of the empirical distributions, confirming that the continuous transport dynamics in the latent space translate to valid biological kinematics in the original CPG space. We discuss this in detail in the supplementary.  

\section{Discussion}
In this work, we transitioned from static epigenetic clocks to continuous, high-dimensional trajectory inference, modeling lifelong human methylation dynamics directly from cross-sectional snapshots. By coupling an age-regularized VAE with the DeepRUOT framework, our pipeline bridges macroscopic population phenomena with microscopic molecular kinematics. However, limitations remain: our multi-decade age discretization introduces a coarse temporal constraint that may smooth over rapid life-stage transitions (e.g., puberty or menopause), while the universal pan-tissue strategy deliberately suppresses localized, tissue-specific metabolic kinetics. Furthermore, while these analyses offer novel insights into normal tissue dynamics, it remains to be explored how these trajectories are rewired in pathological settings, particularly in cancer, where DNA methylation instability frequently drives pathways linked to oncogenic transformation and progression. Furthermore, while simpler baseline methods like PCA capture the linear mean shift of aging, they are structurally insufficient for generative modeling. PCA and standard VAEs lack continuous transport mechanisms, and they completely miss multi-dimensional stochastic noise (epigenetic drift, captured by RUOT's diffusion score $s_\theta$) and non-conservative survivorship bias (captured by the growth field $g_\theta$). High-dimensional inference is required to resolve these non-linear dynamics. Consequently, simulating continuous stochastic dynamics via neural ODEs into regions of expanding late-life variance naturally results in a slight over-extrapolation (flaring) at the low-density extremes of our oldest cohorts (Figure 2b). This marginal stretch is a recognized artifact of modeling outward diffusion without hard spatial boundaries, routinely observed in single-cell trajectory inference, but accurately reflects the biological reality of entropic variance expansion. These boundaries highlight compelling avenues for future research.


\appendix
\clearpage

\section{Supplementary Material}

This supplementary document provides data distributions, and empirical validation figures for epigenetic archetypes supporting the main text. 

\subsection{Dataset Statistics}
Table~\ref{tab:age_distribution} details the stratification of the 12,439 pan-tissue DNA methylation samples across the four discrete chronological age cohorts used as temporal snapshots for trajectory inference.
\begin{table}[H]
    \centering
    \caption{Distribution of DNA methylation samples across chronological age bins.}
    \begin{tabular}{llc}
    \toprule
    Timepoint & Age Range (Years) & Number of Samples \\
    \midrule
    $t_0$ & 0--20 & 3,362 \\
    $t_1$ & 21--40 & 2,041 \\
    $t_2$ & 41--60 & 3,967 \\
    $t_3$ & 61--80 & 3,069 \\
    \midrule
    \textbf{Total} & & \textbf{12,439} \\
    \bottomrule
    \end{tabular}
    \label{tab:age_distribution}
\end{table}

\subsection{Empirical Verification of CpG-Space Trajectories}

To verify that the generated paths accurately reflect real biological progression, we overlaid the decoded continuous mean trajectories onto the raw cross-sectional methylation data. 

\begin{figure}[H]
    \centering
    \includegraphics[width=0.9\linewidth]{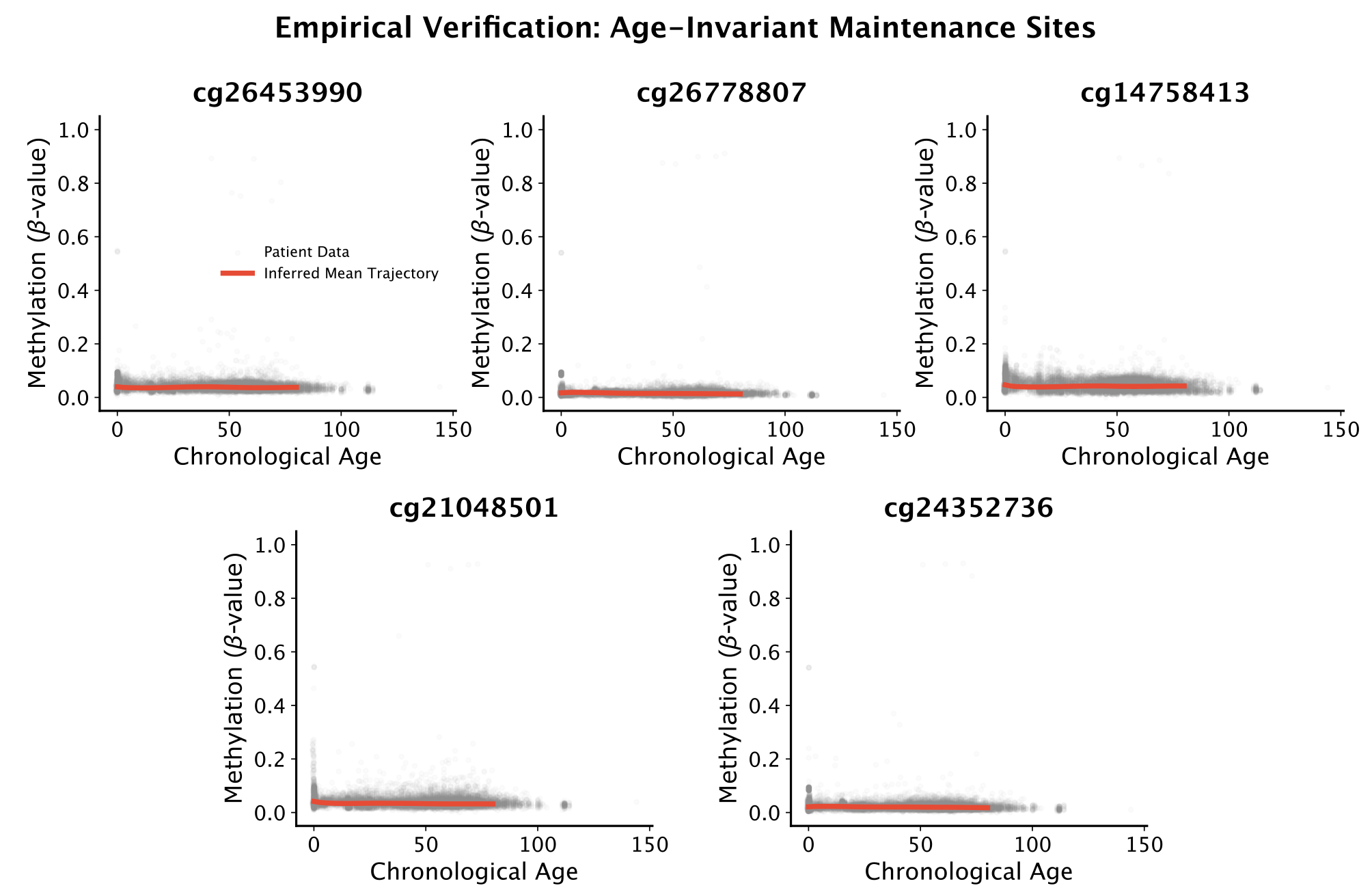}
    \caption{Empirical verification of Age-Invariant Maintenance Sites. The inferred mean continuous trajectory (red line) closely maps the stable baseline across the raw cross-sectional cloud (gray scatter).}
    \label{fig:verify_invariant}
\end{figure}

\begin{figure}
    \centering
    \includegraphics[width=0.9\linewidth]{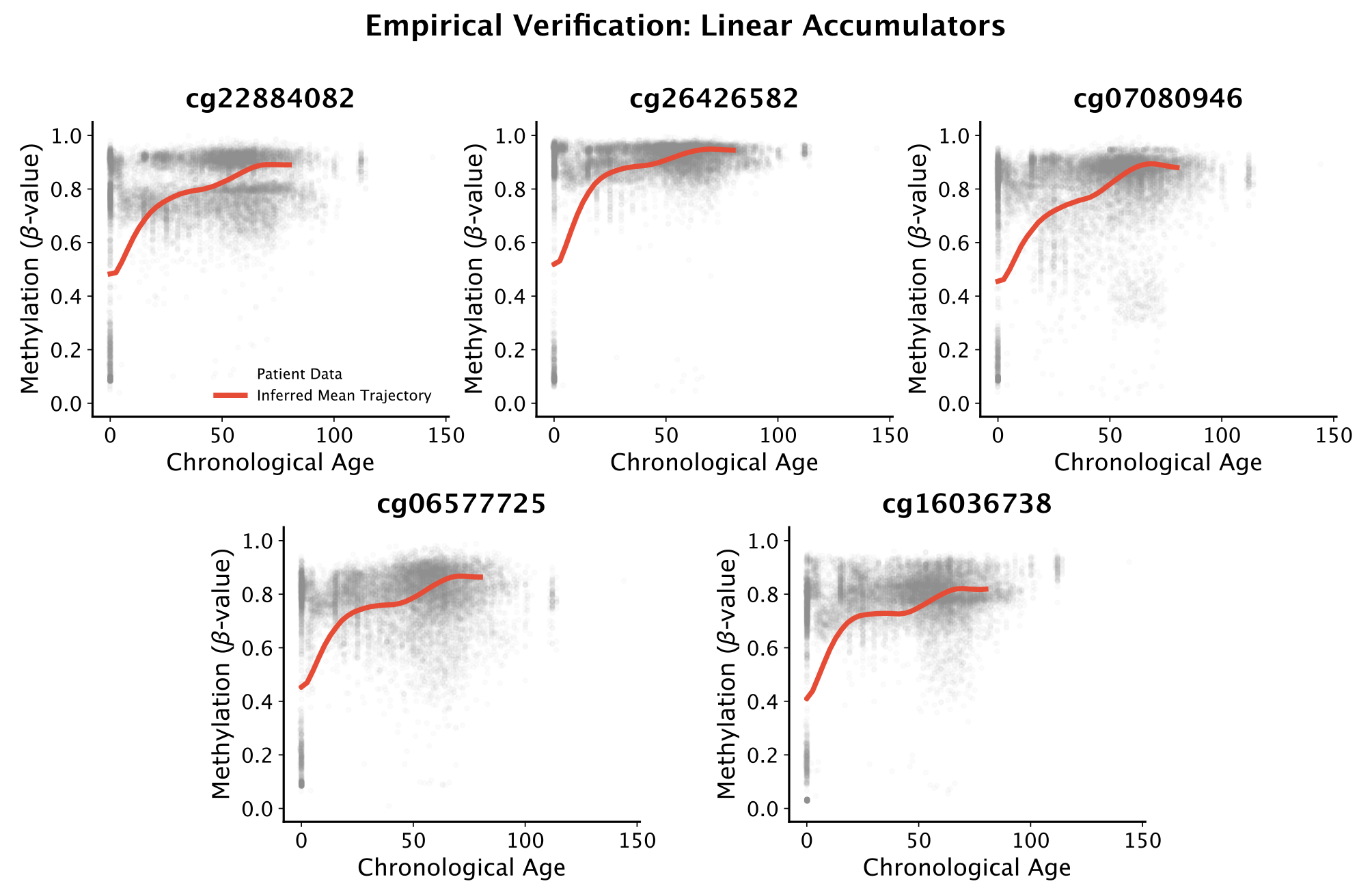}
    \caption{Empirical verification of Linear Accumulators. The continuous trajectory correctly tracks the steady, directional gain in methylation observed across independent profiles.}
    \label{fig:verify_accumulators}
\end{figure}

\begin{figure}
    \centering
    \includegraphics[width=0.9\linewidth]{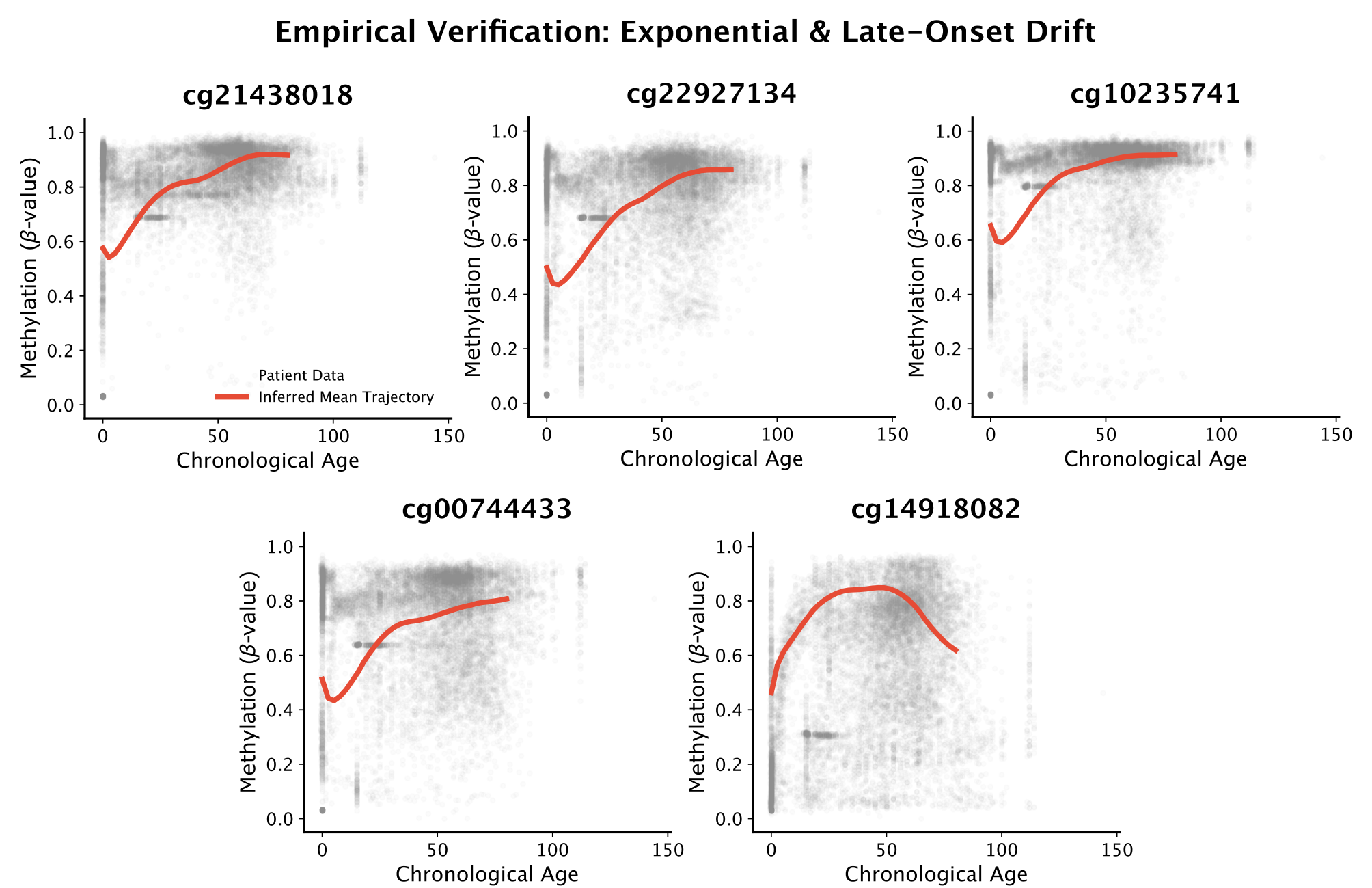}
    \caption{Empirical verification of Exponential \& Late-Onset Drift sites. The model successfully captures complex, non-monotonic, and accelerated late-life dynamics, aligning with the dense distribution of the underlying population.}
    \label{fig:verify_drift}
\end{figure}

For the \textit{Age-Invariant Maintenance Sites} (Supplementary Figure~\ref{fig:verify_invariant}), the model accurately infers a flat, stable baseline, remaining perfectly centered within the dense, unchanging distribution of the raw methylation values across all age bins. For the \textit{Linear Accumulators} (Supplementary Figure~\ref{fig:verify_accumulators}), the inferred paths smoothly capture the steady, monotonic accumulation of methylation, tightly fitting the age-dependent trend of the underlying population from birth through late adulthood. 

Most notably, for the \textit{Exponential \& Late-Onset Drift} markers (Supplementary Figure~\ref{fig:verify_drift}), the model successfully adapts to complex non-linear kinetics. It accurately maps the periods of early-life stability followed by the accelerated, high-variance divergence characteristic of late-life epigenetic entropy. This precise alignment demonstrates that the generative transport process correctly interpolates the shifting density of the population without over-smoothing biologically meaningful stochasticity.

\subsection{Gene Ontology Enrichment}

\begin{figure}
    \centering
    \includegraphics[width=0.9\linewidth]{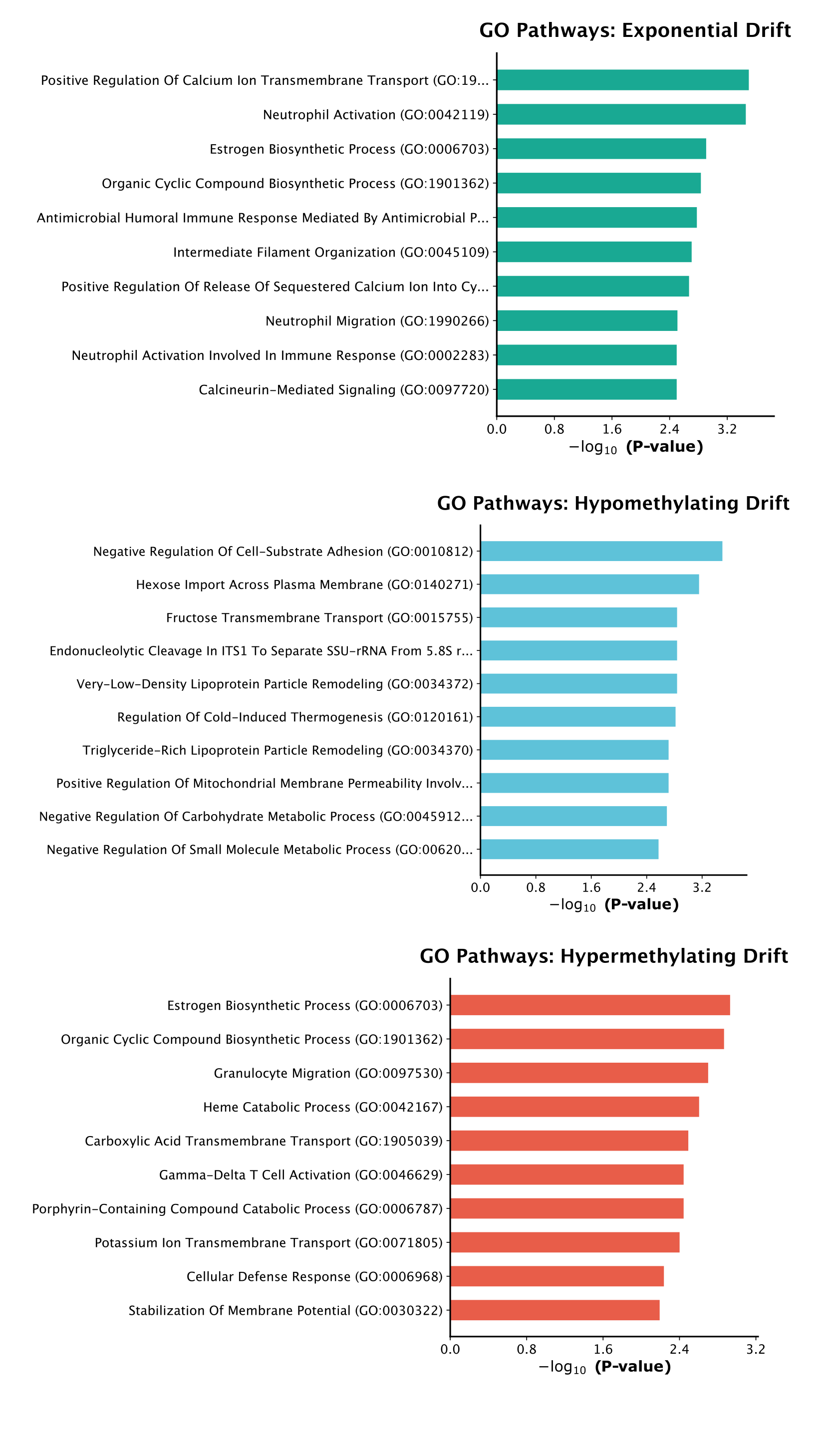}
    \caption{GO Enrichment for Exponential, Hypermethylation and Hypomethylation archetypes}
    \label{fig:go_enrichment}
\end{figure}

To empirically validate archetypes against known biological pathways, we isolated the top 150 most dynamic CpG loci for each of the three primary shifting archetypes (Hypermethylating, Hypomethylating, and Exponential Drift). We subjected these archetype-specific gene sets for biological interpretation using the gene ontology (GO -  Biological Process 2023) database to determine if our mathematically defined clusters map to distinct biological functions. The GO enrichment is show in  figure ~\ref{fig:go_enrichment}. We note that GO enrichment analysis revealed convergent changes in immune/inflammatory signalling, ion transport and calcium-mediated signal transduction, metabolic reprogramming, and lipid/steroid biosynthesis. Across the drift signatures, pathways related to neutrophil activation and migration, cellular defense responses, calcium signalling, nutrient transport, and estrogen metabolism were recurrently enriched, suggesting coordinated remodelling of immune function, cellular signalling, and metabolic state during the drift process. Several of these processes are known correlates of ageing. For instance, immune related adaptations referred to as "Inflammageing" \cite{ferrucci2018inflammageing} results in higher levels of blood inflammatory markers in older subjects. Further, ageing and age-related disease are also known to be associated with deterioration of calcium homeostasis \cite{herraiz2015ageing}, as well as metabolic and mitochondrial dysfunction \cite{amorim2022mitochondrial}. These associations of our top CpG markers with the hallmarks of ageing confirms their validity as key features in our model.

\section{Ablations on finer age bins}

In order to check the sensitivity of our approach, we re-trained and validated the model using finer 10-year age bins (T=1 through T=7). 
\begin{table}[H]
\centering
\caption{Performance metrics at 10-year age bins, evaluating trajectory inference robustness across held-out age bins.}
\label{tab:altumage_results_10}
\begin{tabular}{lcc}
\toprule
$T$ & $W_1$ & TMV \\
\midrule
$T=1$ & $3.0173 \pm 0.0025$ & $0.0819 \pm 1.00\times 10^{-4}$ \\
$T=2$ & $2.6267 \pm 0.0113$ & $0.0370 \pm 2.00\times 10^{-4}$ \\
$T=3$ & $2.7436 \pm 0.0096$ & $0.3786 \pm 5.75\times 10^{-5}$ \\
$T=4$ & $2.8725 \pm 0.0210$ & $0.6796 \pm 4.22\times 10^{-5}$ \\
$T=5$ & $2.6344 \pm 0.0382$ & $0.9983 \pm 1.84\times 10^{-5}$ \\
$T=6$ & $3.1875 \pm 0.0144$ & $0.8696 \pm 3.06\times 10^{-6}$ \\
$T=7$ & $2.9622 \pm 0.0368$ & $0.4360 \pm 3.68\times 10^{-7}$ \\
\bottomrule
\end{tabular}
\end{table}

We found the generative trajectories to be exceptionally robust. As shown in the new evaluation, both the Wasserstein-1 ($W_1$) distance and Total Mass Variation ($TMV$) remain highly stable across all seven intervals, exhibiting remarkably low standard deviations (maximum $W_1$ STD of $0.0382$ and $TMV$ STD of $0.0002$). Furthermore, the absolute $W_1$ distances remain tightly bounded ($2.62$ – $3.18$), proving that our continuous latent dynamics do not destabilize or overfit under finer temporal boundaries. 


\end{document}